\documentclass[%
 reprint,
 amsmath,amssymb,
 aps,
]{revtex4-1}

\usepackage{graphicx}
\usepackage{dcolumn}
\usepackage{bm}
\usepackage{amsfonts}
\usepackage{amssymb}
\usepackage[latin1]{inputenc}
\usepackage{textcomp}
\usepackage{epstopdf}

\begin{document}


\title{Probing coherence and noise tolerance in discrete-time quantum walks: unveiling self-focusing and breathing dynamics}

\author{A. R. C. Buarque and W.S. Dias}
\affiliation{
Instituto de F\'isica, Universidade Federal de Alagoas, 57072-900 Macei\' o, Alagoas, Brazil
}%
\begin{abstract}
The sensitivity of quantum systems to external disturbances is a fundamental problem for the implementation of functional quantum devices, quantum information and computation. Based on remarkable experimental progress in optics and ultra-cold gases, we study the consequences of a short-time (instantaneous) noise while an intensity-dependent phase acquisition is associated with a qubit propagating on $N$-cycle. By employing quantum coherence measures, we report emerging unstable regimes in which hitherto unknown quantum walks arise, such as self-focusing and breathing dynamics. Our results unveil appropriate settings which favor the stable regime, with the asymptotic distribution surviving for weak nonlinearities and disappearing in the thermodynamic limit with $1/N$. The diagram showing the threshold between different regimes reveals the quantum gates close to Pauli-Z as more noise-tolerant. As we move towards the Pauli-X quantum gate, such aptness dramatically decreases and the threshold to self-focusing regime becomes almost unavoidable. Quantum gates close to Hadamard exhibit an unusual aspect, in which an increment of the nonlinear strength can remove the dynamics from self-focusing regime.
\end{abstract}
\maketitle

\section{Introduction}
Quantum walks on a lattice have been indicated as a powerful environment for developing quantum algorithms, as well as a versatile and intuitive framework capable of performing any quantum computation~\cite{PhysRevLett.102.180501,10.1145/380752.380758,PhysRevA.70.022314}. In addition, quantum walks have been shown to be ideal testbed for studying and exploring quantum systems~\cite{PhysRevLett.123.230401,PhysRevLett.111.180503,PhysRevLett.124.050502}. Thus, designing and controlling such quantum processes for the long-time dynamics is a fundamental issue that requires a deep understanding.

Quantum noises are the main obstacle for performance improvement of quantum computing, since their presence can destroy a fundamental component: the delicate quantum state of qubits~\cite{PhysRevA.54.3824,PhysRevLett.81.2594,RevModPhys.89.035002,PhysRevA.94.052325,Campbell2017,PhysRevLett.121.190501,PhysRevLett.121.190501}. Decoherence is a physical phenomenon that typically arises from the interaction
of quantum systems and their environment. In discrete-time quantum walks (DTQWs), the decoherence in the quantum gates drives the walker to a spreading rate quadratically slower in the long time limit~\cite{PhysRevLett.91.130602,PhysRevA.67.032304,doi:10.1142/S0219477505002987}. Broken links, simultaneous measurements of chirality and position, random phases and fluctuations in a given preestablished unitary operation can also  induce the same behavior~\cite{PhysRevA.68.062315,ROMANELLI2005137,PhysRevA.74.012312,PhysRevA.74.022310}. Such results have been corroborated by experimental studies that describe the decoherence inducing a cross-over of quantum dynamics from ballistic to diffusive~\cite{PhysRevLett.104.153602,PhysRevLett.106.180403,PhysRevLett.107.233902}. Sub-ballistic and Anderson localized regimes have been reported for quantum walks with specific irregularities~\cite{PhysRevA.81.062123,PhysRevE.99.022117,PhysRevE.100.032106}. Effects of decoherence on discrete-time quantum walks have been associated with a very fast mixing time and uniform distribution regardless of the initial state of the system and the parity of lattice size~\cite{PhysRevA.67.042315,PhysRevE.81.031113,Venegas-Andraca2012}.

Nonlinear phenomena on DTQWs have also been investigated, in which its source emerges from different frameworks~\cite{Shikano2014,PhysRevA.92.052336,PhysRevA.101.062335,Maeda20183687,PhysRevA.93.022329,PhysRevA.75.062333,PhysRevA.101.023802}. An anomalous slow diffusion has been reported for feed-forward DTQWs, a nonlinear quantum walk in which the coin operator depends on the coin states of the nearest-neighbor sites~\cite{Shikano2014}.  Feed-forward DTQWs have provided the dynamics of a nonlinear Dirac particle, with a description of solitonic behavior and the collisional phenomena between them~\cite{PhysRevA.92.052336}. A modified conditional shift operator displaying a dependence on the local occupation probability shows solitonlike propagation and chaotic regime~\cite{PhysRevA.101.062335}. By using DTQWs which combine zero modes with a particle-
conserving nonlinear relaxation mechanism, a conversion of two zero modes of opposite chirality into an attractor-repeller pair of nonlinear dynamics was reported~\cite{PhysRevA.93.022329}. One of the earliest studies reporting the emergent nonlinear phenomena on DTQWs studied the nonlinear self-phase modulation on the wave function during the walker evolution~\cite{PhysRevA.75.062333}.  Restricted only to Hadamard quantum gates, nondispersive pulses and chaoticlike dynamics has been reported. Detailed study exploring other quantum gates reveal a rich setof dynamical profiles, including the self-trapped quantum walks, a localized regime in which the quantum walker remains localized around its initial position~\cite{PhysRevA.101.023802}. An interesting mathematical treatment on nonlinear DTQWs is described in Ref.~\cite{Maeda20183687}.

Although seems natural to consider linearity in quantum regime, quantum-mechanical systems whose effective evolution is governed by a nonlinear dynamics have been described in both optics~\cite{LEDERER20081} and Bose-Einstein condensates~\cite{RevModPhys.71.463,RevModPhys.78.179}, the same environments where quantum walks have shown a remarkable experimental progress~\cite{PhysRevLett.104.153602,PhysRevLett.107.233902,PhysRevLett.106.180403,PhysRevLett.120.260501,PhysRevLett.121.070402,PhysRevLett.124.050502}. By considering that fault-tolerant architectures are built from the understanding of each possible ingredient, acting simultaneously or not, we studied here the weigth of a noise on the dynamics of a quantum walker within a nonlinear framework. We consider a small and instantaneous amount of decoherence on the qubit distribution, which express a measurement process or other environmental intervention. Nonlinear character is associated an intensity-dependent phase acquisition with a qubit propagating on the lattice, which can represent an emerging third-order nonlinear susceptibility at photonic setups or inter-atomic interactions in ultra-cold atomic systems.  By systematically exploring quantum coherence measures, we report appropriate settings which favor the stable regime. However, unstable regimes unveil DTQWs never once attained, such the self-focusing and breathing dynamics. The stability threshold was investigated by varying the quantum gates, as well as the lattice size, which estimates the quantum behavior in the thermodynamic limit.

\section{Model}

Here, we consider an intensity-dependent (nonlinear) phase acquisition associated with a quantum walker propagating on a 1D lattice of interconnected sites~\cite{PhysRevA.101.023802,PhysRevA.75.062333}. More precisely, the walker consists of a qubit, whose state $|\psi\rangle$ is associated with its position and internal state (chirality), which can be described by spin or polarization states. Thus, $|\psi\rangle$ belongs to a Hilbert space $H=H_c\otimes H_p$, with positions described by the orthonormal basis $\{|n\rangle$: $n \in \mathbb{Z}\}$ spanning the position Hilbert space $H_p$, while the internal state is associated with a two-dimensional Hilbert space $H_c$ spanned by an orthonormal basis $\{|R\rangle=(1,0)^{T}$, $|L\rangle=(0,1)^{T}\}$.

Each step of evolution consists in quantum gates $\hat{C}$ located in the lattice sites which act on the quantum walker and shuffles its internal state, followed by spatial shifts to adjacent sites (left or right) according to its new chirality.  Thus, given a general state written as
\begin{equation}
|\psi(t)\rangle=\sum_{n}\left[a_{n,t}|R\rangle+b_{n,t}|L\rangle\right]\otimes|n\rangle,
\end{equation}
in which amplitudes $a_{n,t}$ and $b_{n,t}$ are complex numbers that satisfy $\sum_{n}(|a_{n,t}|^{2}+|b_{n,t}|^{2})=1$, a single step of dynamical evolution is performed by applying the unitary transformation $|\psi(t+1)\rangle=\hat{U}|\psi(t)\rangle$.  

The standard (linear) protocol regards $\hat{U}=\hat{S}(\hat{C}\otimes I_{p})$. $I_{p}$ describes the identity operator in space of positions and $\hat{C}$ is an arbitrary $SU(2)$ unitary operator given by
\begin{eqnarray}
\hat{C}=\cos(\theta)|R\rangle\langle R|
               &-&\sin(\theta)|R\rangle\langle L| \nonumber \\
              & +&\sin(\theta)|L\rangle\langle R| 
               +\cos(\theta)|L\rangle\langle L|,
\end{eqnarray}
in which the parameter $\theta\in [0,2\pi]$ controls the variance of the probability distribution of the walk. The conditional shift operator $\hat{S}$ then performs $\hat{S}|R\rangle\otimes|n\rangle=|R\rangle\otimes|n+1\rangle$ and $\hat{S}|L\rangle\otimes|n\rangle=|L\rangle\otimes|n-1\rangle$.

In our quantum algorithm, the qubit acquires an intensity-dependent (nonlinear) phase in each step of previous protocol. We consider a quadratic nonlinearity depending on the chirality state, which can represent either a nonlinear optical media in photonic setups or the interactions between atoms for ultra-cold atomic systems. Thus, we add to dynamical evolution protocol ($\hat{U}$) one more operator
\begin{eqnarray}
\hat{K}^{t} &=&\sum_{n}(e^{i2\pi\chi|\psi_{n,R}^{t}|^{2}}|R\rangle\langle R|\nonumber \\ && \hspace{.7cm}+ e^{i2\pi\chi|\psi_{n,L}^{t}|^{2}}|L\rangle\langle L|)\otimes|n\rangle\langle n|,
\label{op:nonlinear}
\end{eqnarray}
such that $\hat{U}(t)=\hat{S}(\hat{C}\otimes I_p)\hat{K}^{t-1}$. The parameter $\chi$ denotes the nonlinear strength of the medium and $\chi=0$ restores the standard (linear) protocol. Furthermore, periodic boundary conditions are assumed on the conditional shift operator
\begin{eqnarray}
\hspace{-.3cm} \hat{S}&=& \sum_{n=1}^{N-1}|n+1\rangle\langle n|\otimes|R\rangle\langle R|        
         + \sum_{n=2}^{N}|n-1\rangle\langle n|\otimes|L \rangle\langle L|\nonumber\\
         & &\hspace{.5cm}+|1\rangle\langle N|\otimes |R\rangle\langle R|+|N\rangle\langle 1|\otimes |L\rangle\langle L|,
\end{eqnarray}
in order to describe the $N$-cycle architecture employed here. The important framework of DTQWs on cycles has been used to display how quantum algorithms can be quadratically faster than its classical correspondent~\cite{10.1145/380752.380758} and how decoherence can be useful in quantum walks~\cite{PhysRevA.67.042315}, for example. Rigorous treatment for noiseless DTQWs on $N$-cycles have proved the long-time average probability distribution of finding the qubit in each site as being uniform on the sites for odd-$N$ and non-uniform for even-$N$~\cite{10.1145/380752.380758,Tregenna_2003}. Thus, we assume odd-$N$ lattices with the initial state of qubit given by

\begin{equation}
|\psi(0)\rangle=\frac{1}{\sqrt{2N}}\sum_{n=1}^{N}(|R\rangle+i|L\rangle)\otimes |n\rangle,
\label{eq:ini.state}
\end{equation}
superposed to a weak noise ($\epsilon=10^{-3}/\sqrt{2N}$). The latter express the interaction with environment, which can represent a measurement process. Hence, we evolve the state of a quantum walker whose initial amplitudes at each site are randomly distributed in the interval $[\frac{1}{\sqrt{2N}}-\epsilon,\frac{1}{\sqrt{2N}}+\epsilon]$, in which a proper normalization is employed to ensure the unitary norm of the resulting distribution. 

\section{Results}

We start following the dynamical evolution protocol described above and computing the quantum coherence, whose rigorous measurement framework has only been developed recently~\cite{PhysRevLett.113.140401,He2017}. Among the advisable measures, we compute the $l1$ norm coherence
\begin{equation}
\mathcal{C}_{l1}(t)=\sum_i \sum_{\small i'\neq i} |\rho_{i,i'}(t)|,
\end{equation}
\begin{figure}[!t]
\centering 
\includegraphics[width=8.1cm]{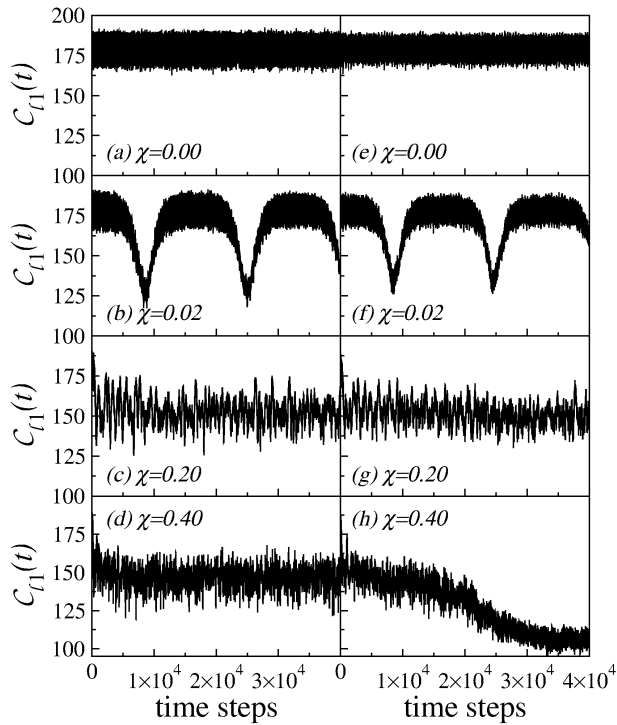}
        \caption{Time evolution of the $l_1$ norm coherence of the whole system for lattices with $N=101$ sites and ruled by quantum gates (a)-(d) $\theta=\pi/4$ and (e)-(h) $\theta=\pi/3$ homegeneously distributed. Oscillatory patterns suggest the lack of stability of the stationary distribution as the nonlinear parameter $\chi$ increases, with the emergence of regular and irregular (chaoticlike) breathing dynamics. Above a critical nonlinear strength, which seems to depend on the quantum gate, the further decrease in quantum coherence after some transient time indicates the wavepacket becoming even narrower.}
        \label{fig1}
\end{figure}
defined as a sum of the absolute values of all off-diagonal elements in the density matrix $\rho=|\psi(t)\rangle\langle\psi(t)|$ under the reference basis. By considering the experimental character an important issue, we observe such quantity being employed to directly measure quantum coherence of an unknown quantum state~\cite{Yuan2020}. Starting from an initial state maximally coherent (eq.~\ref{eq:ini.state})~\cite{PhysRevLett.113.140401,He2017}, we show in Fig.~\ref{fig1} the $l1$ norm coherence of the whole system at each time step for lattices with $N=101$ sites ruled by quantum gates Hadamard ($\theta=\pi/4$) [Figs.~\ref{fig1}(a)-\ref{fig1}(d)] and $\theta=\pi/3$ [Figs.~\ref{fig1}(e)-\ref{fig1}(h)] homogeneously distributed. In absence of nonlinearity ($\chi=0.00$) both quantum gates induce a dynamics with fluctuations over time around a saturation value ($\mathcal{C}_{l1}(t)\sim 2N$), which are fully consistent with previous literature~\cite{He2017}. However, this behavior is heavily modified as $\chi$ grows. In Fig.~\ref{fig1}(b) and Fig.~\ref{fig1}(f), we observe the quantum coherence losing stability and developing regular breaths for a small amount of nonlinearity ($\chi=0.02$). As we further increase the nonlinearity, breathing dynamics gives way to fluctuations whose average value is decreased when compared to the linear regime. Such fluctuations become more rough, suggesting a chaotic aspect. However, both lattices exhibit different behaviors as $\chi$ increases even more. Lattices governed by Hadamard quantum gates remain with coherence exhibiting rough fluctuations around a decreased saturation value ($\mathcal{C}_{l1}(t)\sim 3N/2$), just as $\chi=0.20$. On the other hand, the coherence for lattices ruled by $ \theta=\pi/3 $ quantum gates reveals an additional decrease after an initial transient, with oscillations $\mathcal{C}_{l1}(t)\sim N $ [see Figs.~\ref{fig1}(d) and Figs.~\ref{fig1}(h)].

\begin{figure}[!t]
\centering 
\includegraphics[width=9.0cm]{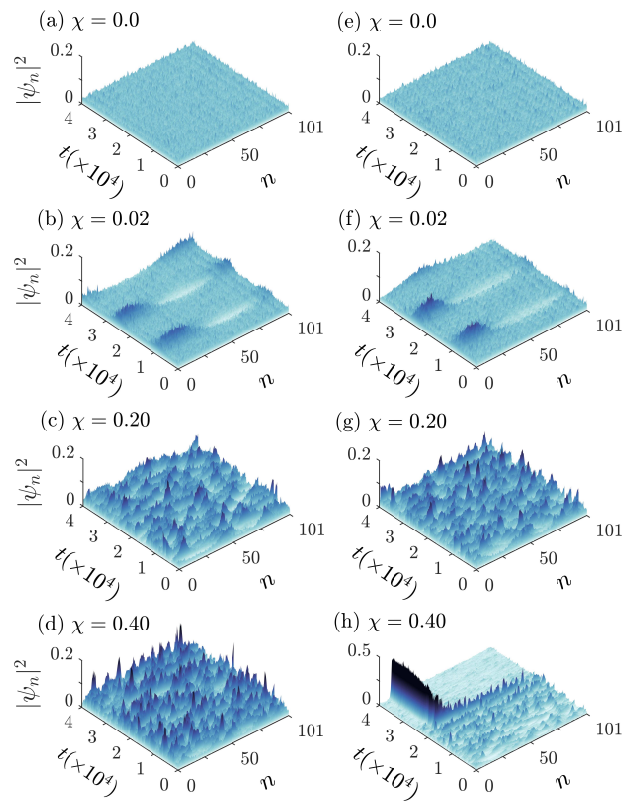}
        \caption{(Color on-line) Time evolution of the density of probability in position space of a quantum walker for the same configurations of $\theta$ and $\chi$ used in Fig.~\ref{fig1}: (a)-(d) $\theta=\pi/4$ and (e)-(h) $\theta=\pi/3$. Corroborating the previous results, we observe clear signatures of regular breathing dynamics for weak nonlinearities. Although both scenarious culminates in a chaoticlike regime as the nonlinear parameter ($\chi$) increases, the self-focusing quantum walk emerges only for $\theta=\pi/3$, which suggests a phenomenology with quantum gate dependence.}
        \label{fig2}
\end{figure}

Since fluctuations in quantum coherence are related to the oscillatory nature of the probability distribution~\cite{PhysRevA.67.042315}, we investigate the probability density distribution $|\psi_n(t)|^2$. We use in Fig.~\ref{fig2} the same configurations shown in Fig.~\ref{fig1}, with Figs.~\ref{fig2}(a-d) and Figs.~\ref{fig2}(e-f) illustrating the dynamical behaviors for lattices governed by $\theta=\pi/4$ and $\theta=\pi/3$ quantum gates, respectively. Corroborating the coherence measures, the spreading of qubit remains uniformly extended over the entire lattice while $\chi=0.0$, which signals the stability of the uniform distribution even after the disturbance. Such stability may disappear when nonlinearity is present, giving way to different regimes. Fully agreeing with the expectations created from the coherence measures, the wavepacket develops regular breathings for very small nonlinearities and irregular breathings, with a chaoticlike aspect, for strong enough nonlinearities. The similarity between both lattices vanishes with $\chi= 0.40$. A self-focusing regime emerging after an initial transient clarifies the strong decreasing of the coherence reported for lattices with $\theta=\pi/3$.

\begin{figure}[!t]
\centering 
\includegraphics[width=5.8cm]{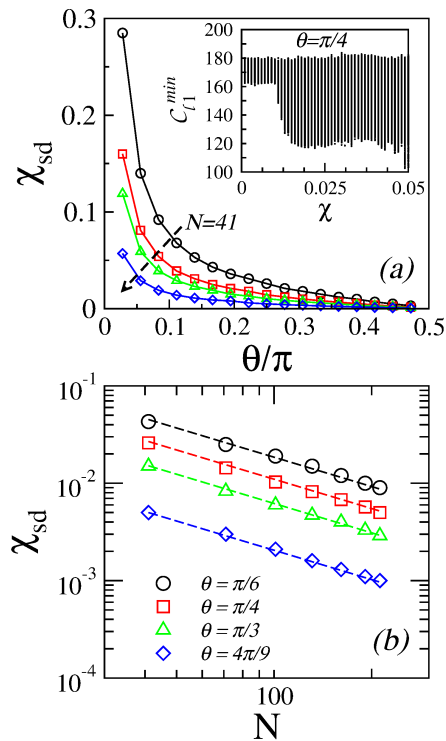}
        \caption{Critical nonlinearity of the stationary state ($\chi_{sd}$) computed for different quantum gates and lattice sizes. (a) $\chi_{sd}$ vs $\theta$ ($\pi$ units) shows quantum gates close to Pauli-Z better able to sustain the stationary regime. A monotonic decreasing of $\chi_{sd} $ as we move $\theta$ towards the Pauli-X gate is also observed, regardless of the size $ N $ (arrow points to the growth of $N$). Minima of the coherence ($\mathcal{C}_{l1}^{min}$) for a Hadamard lattice illustrate the critical point $\chi_{sd}$, above which the distribution develops regular breathings. (b) The $\chi_{sd}$ vs $N$  analysis confirms the size-dependence of the critical nonlinearities and reveals a gate-independent scaling $\chi_{sd} \propto 1/N$.}
        \label{fig3}
\end{figure}

In order to better understand, we follow the time-evolution for a long-time looking for $\chi$ configurations able to remove the qubit dynamics from the limiting distribution. In Fig.~3a we show the relationship between the quantum gates ($\theta $) and the critical nonlinearity $(\chi_{sd})$, above which the distribution becomes unstable. We observe the stationary regime surviving to a greater range of nonlinearities when the system is managed by quantum gates close to Pauli-Z ($\theta=0$). On the other hand, systems governed by quantum gates nearby to Pauli-X ($\theta=\pi/2$) are more sensitive, with minor critical nonlinearities. Thus, quantum gates close to Pauli-Z are less liable to provide qubits with logical errors. The stationary regime threshold exhibits a monotonic decreasing of the critical nonlinearity as we move towards the Pauli-X quantum gates. Such scenario clearly demonstrate the sensitivity associated with the interference terms of the quantum gates, which has its influence amplified by the nonlinear component. The figure also shows a relationship between the critical points and the size $N$, with the arrow pointing to the growth of $N$. The monotonic decline of $\chi_{sd}$ as we left systems ruled by Pauli-Z towards Pauli-X systems remains unchanged. However, the increase of lattice size makes the system more susceptible to instability, since the range of nonlinearities capable of sustaining the stationary regime decreases. The inset illustrate the critical point $\chi_{sd}$, above which the distribution becomes unstable. We recorded minima of the $l1$ norm coherence ($\mathcal{C}_{l1}^{min}$) for every time steps throughout the dynamic evolution of a system ruled by Hadamard quantum gates. By plotting $\mathcal{C}_{l1}^{min}$ as a function of the nonlinear strength $\chi$, we report a well-defined discontinuity which is consistent with the development of regular breaths shown in the previous figures. In Fig.~\ref{fig3}b, we characterize the critical nonlinearity scaling with $1/N$ regardless of quantum gates employed, in close analogy to the dynamics of single electron wavepackets obeying a continuous-time Schr\"odinger equation under similar conditions~\cite{COPIE2020100037,DIAS2019121909}. Such behavior indicates the stationary regime disappearing in the thermodynamic limit ($N\rightarrow\infty$) for any finite nonlinear strength.

\begin{figure}[!t]
\centering 
\includegraphics[width=8.2cm]{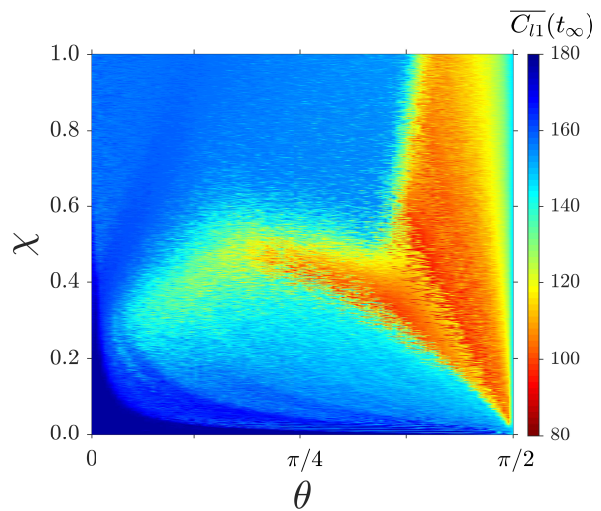}
        \caption{(Color on-line) 
Plot of $\chi$ versus $\theta$ for the long-time average of the $l_1$ norm coherence. Quantum gates close to Pauli-Z are more conducive to sustain the stationary distribution. As we increase $\theta$ towards the Pauli-Z gate ($\theta=\pi/2$), such propensity vanishes and different scenarios emerge as we change $\chi$: breathing dynamics, chaoticlike and self-focusing quantum walks. The last ones are predominant when $\theta$ gets close to Pauli-Z gate. An unusual aspect is reported for quantum gates close to Hadamard, in which an increment of $\chi$ can remove the dynamics from the self-focusing regime. }
        \label{fig4}
\end{figure}

By considering the implementation of a universal set of quantum gates as crucial for a quantum computing architecture, we extend our numerical experiments and show a $\chi~vs~\theta$ diagram in Fig.~\ref{fig4}. Here, we compute the long-time average of $\overline{C}_{l1}(t_\infty)$ for $N=101$ lattice sites. Data reveal the stationary regime surviving for weaker nonlinearities, prevailing for systems configured next to Pauli-Z quantum gate, which is characterized by leaving the basis state $|L\rangle$ unchanged and takes $|R\rangle$ to $-|R\rangle$. Although the Fig.~\ref{fig2} suggests the regimes stationary, breathing, chaoticlike, and self-focusing in ascending order of nonlinearity, the breathing regime persists for systems with quantum gates close to Pauli-Z even for strong nonlinearities. On the other hand, systems with quantum gates nearby to Pauli-X exhibit a fairly narrow range of $\chi$ for breathing dynamics. The emergence of chaoticlike regime is predominantly surrounding the ($\theta=\pi/6$) quantum gates, which arises for an intermediate nonlinear strength. Self-focusing regime appears around the quantum gates of Hadamard, preceded by chaoticlike, breathing and stationary regimes in decreasing order of nonlinearity. However, we observe an unusual threshold between the self-focusing and chaoticlike regimes: The increment on the nonlinear parameter is able to direct the system from the self-focusing to the chaoticlike regime, which persists as the nonlinearity increases. Such remarkable feature agrees with report of self-trapped quantum walks described for narrow qubits spreading in noiseless systems~\cite{PhysRevA.101.023802}. For systems ruled by quantum gates close to Pauli-X, which is characterized by mapping $|L\rangle$ to $|R\rangle$ and $|R\rangle$ to $|L\rangle$, the self-focusing regime emerges for very weak nonlinearities and remains for the strongest nonlinearities. On the other hand, the self-focusing regime is absent as $\theta$ gets very closer to Pauli-Z. 

\section{Conclusions}

In summary, we have implemented a quantum protocol in order to rate the sensitivity to a short-time (instantaneous) noise while nonlinear components are present in a discrete-time quantum walk. The intensity-dependent nonlinearity is based on possible third-order nonlinear susceptibility in optical setups or emergent inter-atomic interactions in ultra-cold atomic systems, while the noise can represent a measurement process or other environmental intervention. Our results unveil optimal sets of operating parameters that favor a stable operation. Quantum gates close to Pauli-Z as more noise-tolerant, contrary to the behavior exhibited by quantum gates nearby Pauli-X gates. When it loses stability, the system may present breathing dynamics and self-focusing, hitherto unknown quantum walks. The crossover from the uniform distribution to unstable regime decaying as $1/N$ reveals a fault-intolerant system in the thermodynamic limit ($N\rightarrow \infty$), i. e., nonlinearities may be responsible for the unable to encode and decode qubits robustly. In addition to contributing to the deeper fundamental understanding on discrete-time quantum walks, breathing and self-focusing quantum walks also bring applicability prospects as for  microresonators, lensing-like effects and wave guiding, which arises from accumulated self-focusing.

\section{Acknowledgments}

This work was partially supported by CNPq (The Brazilian National Council for Scientific and Technological Development), CAPES (Federal Brazilian Agency) and FAPEAL (Alagoas State Agency).

\bibliographystyle{nature}
\bibliography{referencias}

\end{document}